%% file: main.tex
\documentclass[sigconf]{acmart}
\AtBeginDocument{%
  \providecommand\BibTeX{{%
    \normalfont B\kern-0.5em{\scshape i\kern-0.25em b}\kern-0.8em\TeX}}}

\usepackage{subcaption}
\usepackage{float}
\usepackage[english]{babel}

\newcommand{\CoTwo}{CO\textsubscript{2}}

\addto\extrasenglish{%
}
\renewcommand\footnotetextcopyrightpermission[1]{} 		%
\setcopyright{none}

\begin{document}

\title{A User Interface Study on Sustainable City Trip Recommendations}

\author{Ashmi Banerjee}
\authornote{Both authors contributed equally to this research.}
\email{ashmi.banerjee@tum.de}
\affiliation{%
  \institution{Technical University of Munich}
  \streetaddress{Boltzmannstrasse 3}
  \city{Garching}
  \country{Germany}
  \postcode{85748}
}
\author{Tunar Mahmudov}
\authornotemark[1]
\email{tunar.mahmudov@tum.de}
\affiliation{%
  \institution{Technical University of Munich}
  \streetaddress{Boltzmannstrasse 3}
  \city{Garching}
  \country{Germany}
  \postcode{85748}
}
\author{Wolfgang Wörndl}
\email{woerndl@in.tum.de}

\affiliation{%
  \institution{Technical University of Munich}
  \streetaddress{Boltzmannstrasse 3}
  \city{Garching}
  \country{Germany}
  \postcode{85748}
}

\renewcommand{\shortauthors}{Banerjee, et al.}

\begin{abstract}
The importance of promoting sustainable and environmentally responsible practices is becoming increasingly recognized in all domains, including tourism. The impact of tourism extends beyond its immediate stakeholders and affects passive participants such as the environment, local businesses, and residents. City trips, in particular, offer significant opportunities to encourage sustainable tourism practices by directing travelers towards destinations that minimize environmental impact while providing enriching experiences. This is where Tourism Recommender Systems (TRS) can play a critical role. By integrating sustainability features in TRS, travelers can be guided towards destinations that meet their preferences and align with sustainability objectives.

This paper aims to investigate how different user interface design elements affect the promotion of sustainable city trip choices. We explore the impact of various features on user decisions, including sustainability labels for transportation modes and their emissions, popularity indicators for destinations, seasonality labels reflecting crowd levels for specific months, and an overall sustainability composite score. Through a user study involving mockups, participants evaluated the helpfulness of these features in guiding them toward more sustainable travel options.

Our findings indicate that sustainability labels significantly influence users towards lower-carbon footprint options, while popularity and seasonality indicators guide users to less crowded and more seasonally appropriate destinations. 
This study emphasizes the importance of providing clear and informative sustainability information to users, which can help them make more sustainable travel choices. It lays the groundwork for future applications that are capable of recommending sustainable destinations in real time.

\end{abstract}


\maketitle
\pagestyle{plain} %
\input{sections/0_intro.tex}

\input{sections/3_related.tex}
\input{sections/4_methodology.tex}
\input{sections/5_results.tex}

\input{sections/6_conclusion.tex}

\bibliographystyle{ACM-Reference-Format}
\bibliography{references}

\end{document}

%% file: sections/0_intro.tex
\section{Introduction}

The importance of advocating for sustainable and environmentally responsible practices is gaining popularity across various sectors, including tourism. The impact of tourism extends beyond direct stakeholders, affecting non-participating entities such as the environment, local businesses, and residents. 
The relationship between tourism and the environment is intricate, encompassing activities that yield both positive and negative outcomes. While tourism can contribute to environmental protection, conservation efforts, and heightened awareness of ecological values, it can also have adverse effects, such as contributing to climate change, resource depletion, and the issues of overtourism or undertourism~\cite{camarda2003environmental,gossling2017tourism}. 
Moreover, the intricacies within the tourism domain are compounded by factors such as seasonality, travel regulations, and resource constraints like availability of airline tickets and hotel accommodations~\cite{balakrishnan2021multistakeholder}. 
This is where Tourism Recommender Systems (TRS) play a pivotal role, aiding in trip planning by offering personalized recommendations for accommodations, activities, destinations, and more, all while considering the inherent constraints of the domain~\cite{ISINKAYE2015261}.

A well-designed TRS can guide travelers towards destinations that align with their preferences and meet sustainability objectives, thereby addressing the challenges of overtourism and undertourism. 
Overtourism refers to situations where destinations become overwhelmed with excessive visitors, often exacerbated by factors such as the proliferation of low-cost aviation, affordable transportation options, social media influence, and platforms like Airbnb~\footnote{https://www.airbnb.com/}. Conversely, undertourism occurs when destinations remain underexplored due to inadequate infrastructure, limited publicity, and poor accessibility~\cite{Gowreesunkar2020}. Both phenomena yield detrimental effects. Overtourism endangers the preservation of historic city centers, adversely impacts the environment, disrupts residents' lives, and diminishes tourists' experiences, leading to challenges such as environmental pollution and inflated housing prices in affected cities~\cite{dastgerdi2023post}. 
Conversely, the absence of tourists or undertourism also negatively affects the tourism and hotel industries in lesser-known destinations~\cite{camarda2003environmental, Gowreesunkar2020}.

To address these challenges, TRS should be designed to consider the needs and interests of all stakeholders, promote sustainable tourism practices, and foster responsible tourism behavior among users. 
While traditional fairness considerations address the needs of direct stakeholders like travelers and service providers, a fair TRS should extend this scope to include non-participating stakeholders or \textit{society} affected by tourism activities. 
\textit{Society}, in this context, encompasses non-participating stakeholders who may experience repercussions such as elevated housing costs, environmental degradation, and traffic congestion due to heightened tourism activities in the region~\cite{banerjee2023review,banik2023understanding,banerjee2024sf,banerjee2023fairness}.

In this paper, we examine the influence of various user interface design components on promoting sustainable choices for city trip recommendations.
We analyze the effects of several features, such as sustainability labels for transportation modes and associated emissions, popularity indicators for destinations, seasonality labels indicating crowd levels for specific months, and a composite score representing overall sustainability on user decision-making.
Through a user study, we validate our future application design by soliciting feedback from users regarding the effectiveness of these indicators in capturing sustainability and aiding their decision-making process when selecting a sustainable destination. 
To this end, our paper makes the following contributions:
\begin{itemize}
    \item Assessing the effectiveness of the visualizations in influencing users' travel choices by determining whether they successfully encouraged users to prioritize sustainability.
    \item Gathering insights into user preferences and feedback.
    \item Evaluating our mockup to lay the groundwork for a future application capable of recommending sustainable destinations in real time.
\end{itemize}

Our paper is structured as follows--- ~\autoref{section: related} reviews prior research in this area,~\autoref{section: methodology} outlines our proposed methods for the user study,~\autoref{section: results} presents the outcome and finally~\autoref{section: conclusion} concludes the paper by discussing limitations and suggesting avenues for future work.

%% file: sections/3_related.tex
\section{Related Work} \label{section: related}

Studies have explored TRS from numerous angles, targeting consumers' and providers' needs and goals. 
Beyond well-known platforms like Tripadvisor and Booking.com,~\citet{alrasheed2020multi} propose a multi-layer TRS that gathers user preferences to provide tailored destination recommendations, incorporating factors like travel details, budget, and user preferences and presents a list of potential destinations favored by similar users.
Subsequently,~\citet{batet2012turist} introduce a model, Turist@, which provides personalized recommendations for cultural and leisure activities at the traveler's destination by initially gathering user preferences through a survey and continuously refining them based on user interactions within the system.

Related work has also explored sustainable travel and recommended environmentally friendly destinations.
For instance, platforms such as \citet{Fairbnb} facilitate ethical and sustainable tourism by allocating 50\% of its fees to local projects chosen by travelers.
Similarly, \citet{Ecobnb} distinguishes itself by linking travelers with sustainable accommodation choices and enforcing strict environmental standards, including renewable energy, organic food, water conservation, and many more, to minimize travelers' ecological impact. 
Furthermore, \citet{merinov2023sustainability} investigates the adverse impacts of user-centric tourism. They introduce a multistakeholder approach that addresses the issue of managing tourist traffic to safeguard popular sites from overtourism while encouraging the growth of lesser-known destinations by evenly spreading tourists across various locations.
\citet{banik2023understanding} address the benefits of integrating sustainable recommendations in TRS, illustrating how these can encourage tourists to consider sustainable and less popular destinations, thereby addressing overtourism and undertourism. 
On the other hand,~\citet{noubari2023dynamic} created the "Destination Finder" application, which lets users interactively refine preferences with a color-coded map. Their study confirmed its effectiveness in addressing visual challenges and highlighted the importance of user feedback and interface design.
Our research stands out from previous studies by evaluating visualization strategies for different factors like \CoTwo~emissions, destination popularity, and monthly crowdedness estimations to offer sustainable destination recommendations. Additionally, we also examine understanding user preferences and decision-making tradeoffs.

%% file: sections/4_methodology.tex
\section{Methodology} \label{section: methodology}

Recent studies have explored the concept of Societal Fairness (\textit{S-Fairness}) in TRS~\cite{banerjee2023review,banik2023understanding,banerjee2024sf,banerjee2023fairness}. This concept focuses on the impact of tourism on non-participating stakeholders, particularly residents. It addresses concerns such as increased housing prices, environmental pollution, and traffic congestion resulting from heightened tourist activities in the area~\cite{banerjee2023fairness}. 
In this paper, we aim to be fair to society by recommending sustainable destinations to the users.
Our approach adopts the modeling framework proposed by~\citet{banerjee2024sf}, which integrates various components —-- the \textit{emission tradeoff index}, \textit{popularity index}, and \textit{seasonal demand index} to establish an overall \textit{S-Fairness indicator} for each month to cities accessible from users' starting points. 
A lower indicator value indicates more appeal in sustainability. 
This metric is helpful for the following:
\begin{itemize}
    \item Encouraging environmentally friendly options by recommending destinations with lower \CoTwo~emissions, thus promoting sustainable travel practices.
    \item Directing tourists to lesser-known yet appealing destinations, alleviating pressure on over-visited areas, and diversifying tourist traffic.
    \item Balancing visitor numbers by selecting destinations in their low season, mitigating overtourism during peak times, and preventing undertourism during off-peak periods.
\end{itemize}

To gain deeper insights into user decision-making processes when choosing their next destination for vacation, we conducted a user study where participants were presented with mockups of various user interfaces. These interfaces highlighted the individual components of the \textit{S-Fairness indicator} and the overall scores.
Detailed descriptions of the user interface designs presented are outlined in~\autoref{subsection: ui_design}, while~\autoref{subsection: survey_setup} elucidates the methodology for the user study.

\begin{figure*}[htbp]
    \centering
    \begin{subfigure}{0.8\textwidth}
        \centering
        \includegraphics[width=\textwidth]{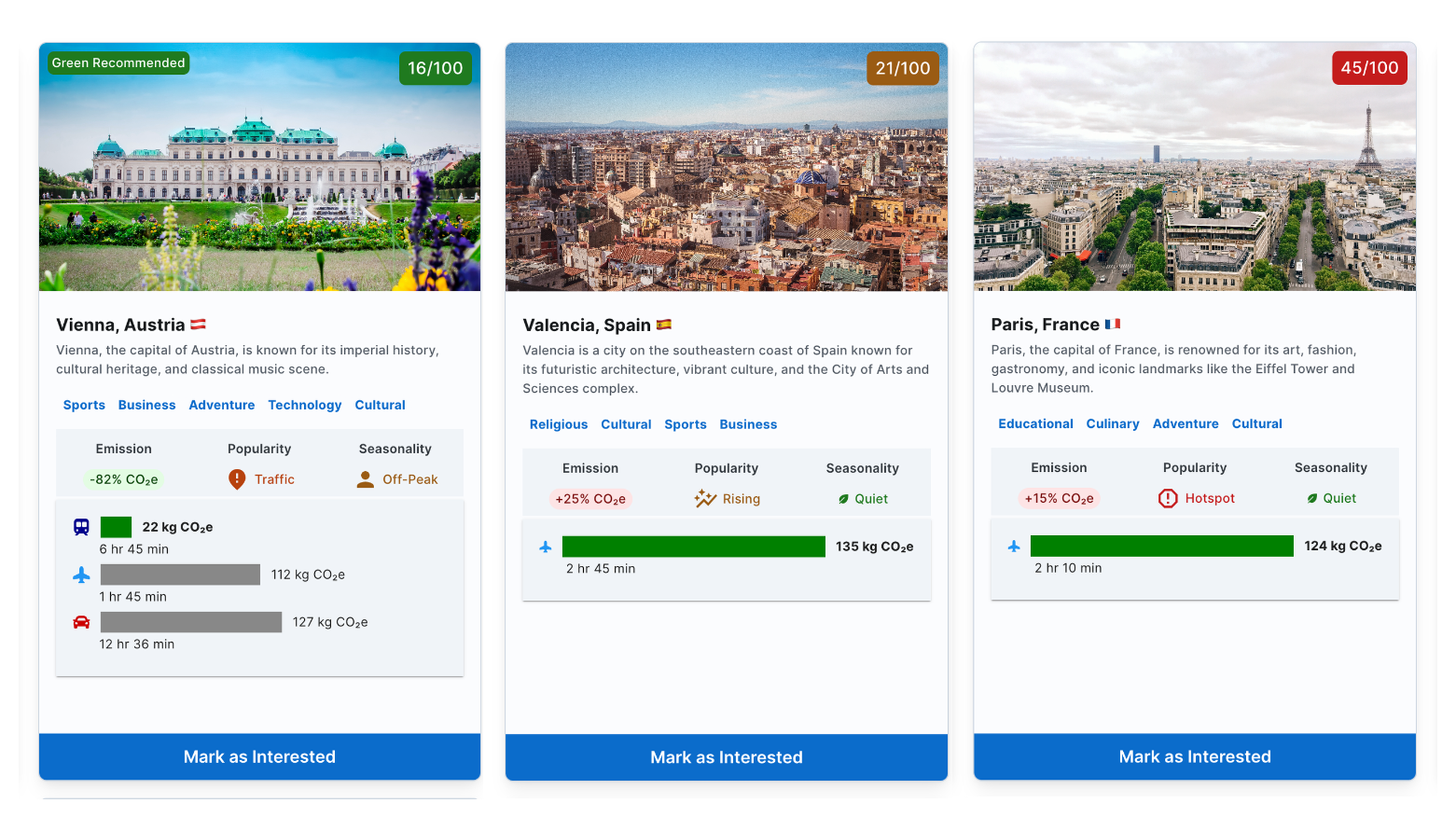}
        \caption{Card items that were shown to the users}
        \label{fig: card_item}
    \end{subfigure}
    \hfill
    \begin{subfigure}{0.6\textwidth}
        \centering
        \includegraphics[width=\textwidth]{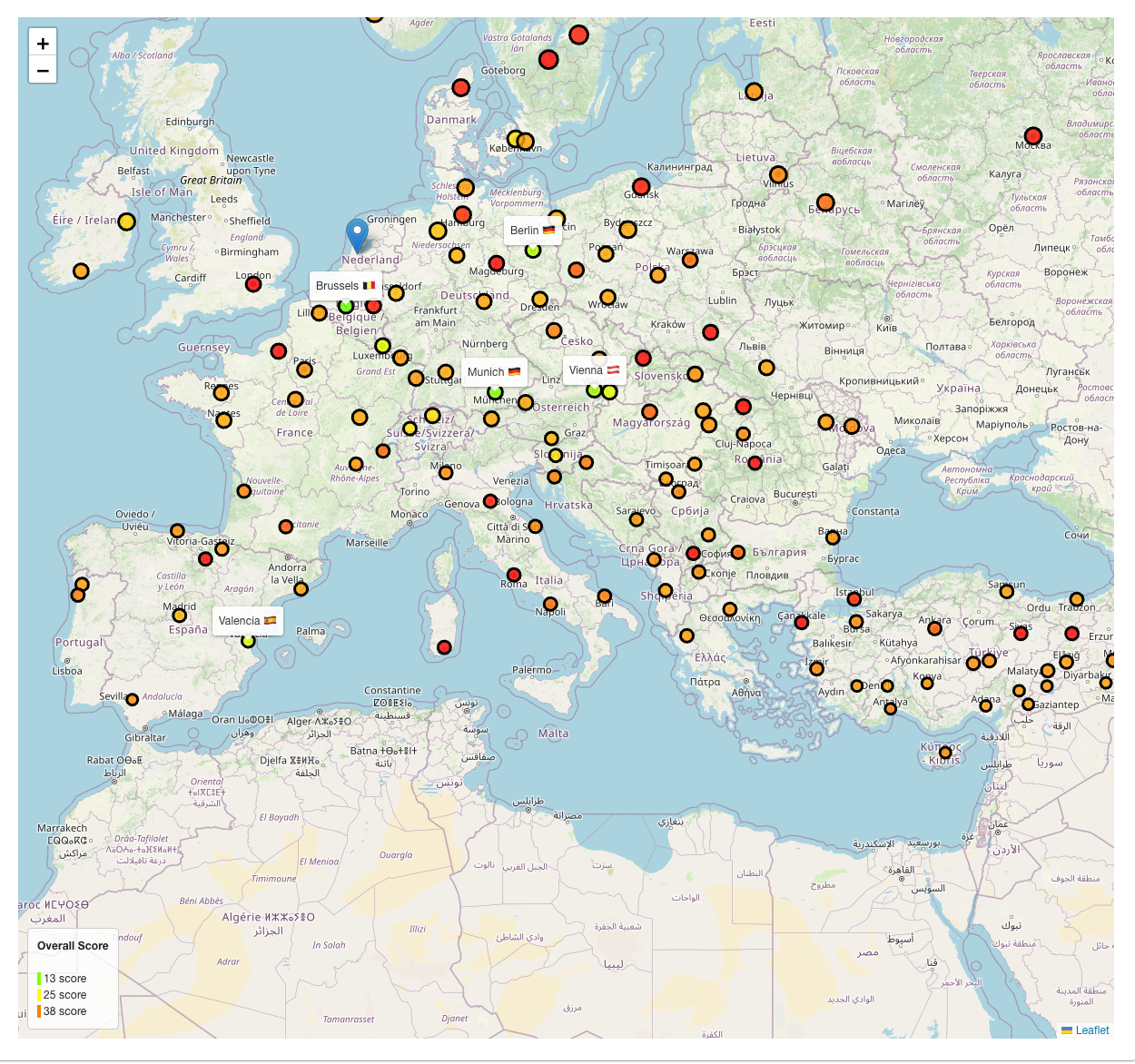}
        \caption{Map view}
        \label{fig: map_view}
    \end{subfigure}
    \caption{Views shown to the users}
    \label{fig: card-map-views}
\end{figure*}

\subsection{User Interface Design} \label{subsection: ui_design}
Participants were asked to consider planning city trips in Europe, starting from a chosen location for a specific month.
The interface displayed potential European destinations for a specific month in card and map formats, as depicted in~\autoref{fig: card-map-views}.
The purpose was to evaluate the effectiveness of different sustainability component visualizations.
Throughout the mockups, we followed a traffic light color scheme~\cite{uxmag_traffic_light_colours}.
We used green for the most sustainable options, amber for neutral choices, and red for the least sustainable alternatives, allowing users to quickly understand the sustainability levels of different components within the application.

\subsubsection{Card View}

In this layout, users were presented with a list of three cities, accompanied by their respective countries, depicted with flag icons, representative photos, and brief descriptions of their attractions (\autoref{fig: card_item}). The objective of this layout was to break down the visualization of individual components of the \textit{SF indicator} and assess the effectiveness of the representations of various labels used to visualize these components.

We incorporated travel logistics, indicating travel time from the starting city, estimated duration, and  \CoTwo~emissions for different modes of transport --- trains, driving, and flight. 
We also displayed emission tradeoffs from the user's starting point and ranked transportation modes by their \CoTwo~emissions, with lower values suggesting a more sustainable choice. 
Cities with higher emissions than average had their emission tradeoff index labeled in red, while those with lower emissions were labeled in green.
In addition, we used labels like \textit{"rare find}, \textit{"traffic"}, \textit{"rising"}, or \textit{"hotspot"} to indicate the popularity level of each destination as low, medium, or high, respectively. 
The goal was to encourage users to opt for less popular yet appealing destinations, fostering a more balanced and sustainable tourism experience. 
Furthermore, color-coded labels such as \textit{"quiet"}, \textit{"off-peak"},  \textit{"busy"}, or \textit{"crowded"} were used to represent the seasonality of each city in the selected month.
This approach aims to alleviate destination overcrowding by promoting travel during months with lower seasonal demand.

Each card also displayed an overall sustainability score (\textit{SF indicator}) at the top right corner. This indicator was calculated as a weighted sum, considering emissions relative to the user's starting point, destination popularity, and crowd levels (seasonal demand) for the selected month~\cite{banerjee2024sf}. A lower score indicated a more sustainable destination. 
Additionally, the most sustainable option was labeled as \textit{"green recommended"}, \textit{"lowest emission"}, \textit{"hidden gem"}, or \textit{"least crowded"} to encourage users to choose it. 
The cards were sorted in ascending order of the overall score, and the scores were color-coded to assist users in understanding the sustainability of their choices.

\subsubsection{Map View}
We also provided users with a map view (\autoref{fig: map_view}) to enhance their decision-making process by offering a visually intuitive way to explore and select sustainable destinations. 
In this view, the \textit{SF indicators} for reachable destinations from the user's selected starting point were color-coded, with green indicating the most sustainable options and red representing the least sustainable ones. The starting point was marked with a blue location icon. 
While the card view offered a nuanced breakdown of individual components of sustainability, the map view was constructed to provide a more high-level color-coded overview of the sustainable destinations reachable from the user's starting point.

\subsection{Survey Setup} \label{subsection: survey_setup}

The survey instrument was created utilizing Qualtrics Experience Management Software~\footnote{https://www.qualtrics.com}, an online survey platform. Participants were recruited via the online crowdsourcing platform Prolific~\footnote{https://www.prolific.com}, known for its effectiveness in subject recruitment for academic research. 
Targeting European individuals who listed travel as one of their interests, the questionnaire was distributed in English using Prolific's advanced pre-screening features, resulting in 200 completed responses. 
We aimed for equal representation, with 50\% male and 50\% female participants in the preset distribution to ensure gender diversity. 
To safeguard participants' privacy, demographic questions regarding age, gender, and nationality were excluded. 
The questionnaire predominantly featured Likert scale questions~\cite{jamieson2004Likert} to elicit detailed user feedback and insights. This approach was chosen to ensure a thorough understanding of user interactions with the mockup, laying a foundation for future enhancements to support sustainable travel planning applications.

%% file: sections/5_results.tex
\section{Experimental Evaluation} \label{section: results}

Participants were shown mockups and asked to assess the effectiveness of each color-coded label in aiding their selection of sustainable alternatives. Additionally, they evaluated the overall presentation formats, including the card and the map view. 
This section investigates the details of the results of our study.

\input{sections/results/3_1_results.tex}

\input{sections/results/3_2_results.tex}

%% file: sections/results/3_1_results.tex
\subsection{Helpfulness of different sustainability labels}

\begin{figure*}[htbp]
    \centering
    \includegraphics[width=0.75\textwidth]{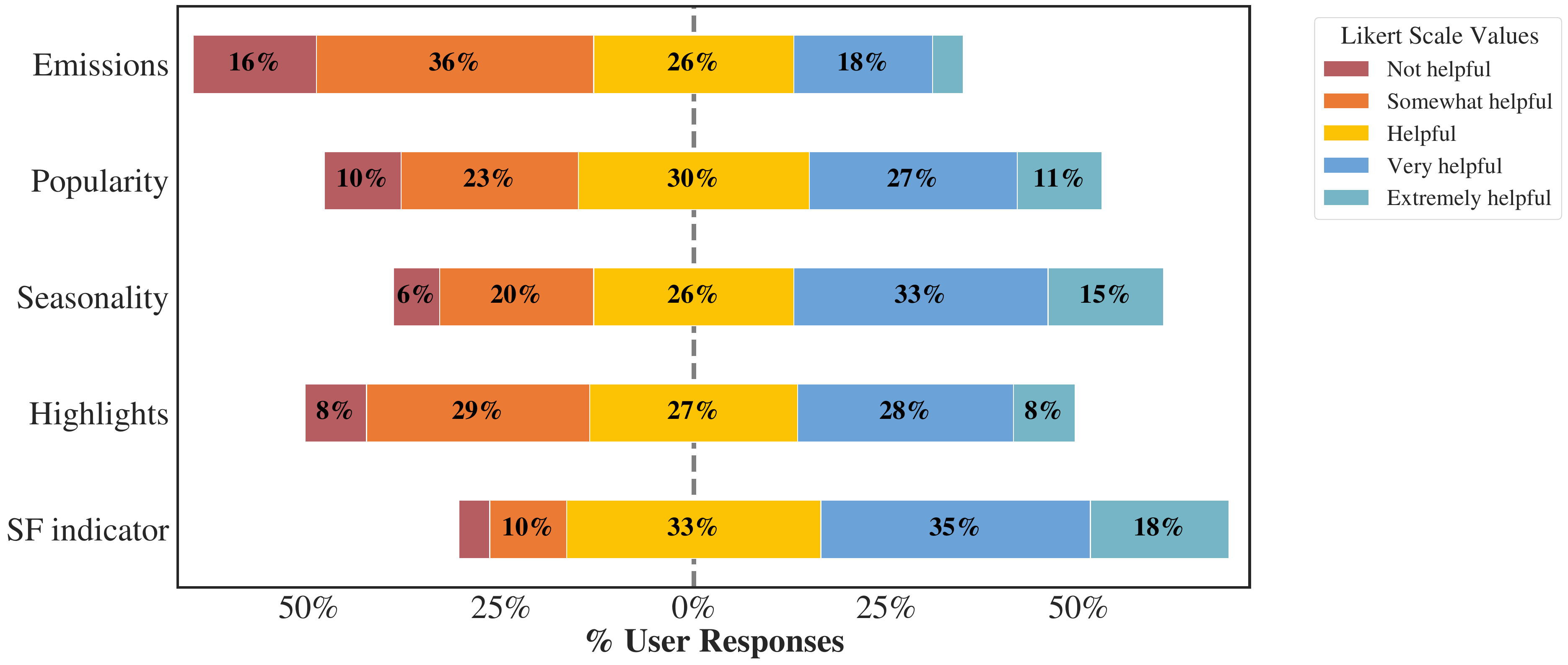}
    \caption{Effect of the different user interface labels on users' decision-making.}
    \label{fig: ui_labels_eval}
\end{figure*}

The evaluation of user responses, as illustrated in~\autoref{fig: ui_labels_eval}, highlights the influence of various user interface labels on decision-making regarding sustainable city trip recommendations. 
Participants provided ratings on Likert Scale statements~\cite{joshi2015likert}, spanning from \textit{"not helpful"} to \textit{"extremely helpful"}, using a scale of 1 to 5. These ratings gauged their agreement levels regarding various sustainability aspects, including emissions, popularity, seasonality, and the effectiveness of highlighting labels and the color-coded scheme for the overall \textit{SF indicator}.

\subsubsection{The highlighting of transportation modes and their emission labels influencing their decision to select a transportation option with a lower \CoTwo~emission value}

The mockup displayed a sample of three reachable destinations from the user's starting location, using three transportation modes --- driving, train, and flights sorted by their respective \CoTwo~emission values. 
Users were asked if organizing transportation choices by emissions influences their selection of destinations with a smaller carbon footprint for travel. 
They rated the presentation of transportation options and associated \CoTwo~emissions as seen in~\autoref{fig: ui_labels_eval}. 
The largest group of 36\% of participants, totaling 72 individuals, found these labels as \textit{"somewhat helpful"}. The second-largest group, comprising 26\%, found them to be \textit{"helpful"}, while 18\% thought they were \textit{"very helpful"}. Notably, only 4\% reported being \textit{"extremely helpful"}, and 16\% of the participants reported being \textit{"not helpful"}. 
Although \textit{"somewhat helpful"} was the predominant response in~\autoref{fig: ui_labels_eval}, the average response leaned towards \textit{"helpful"}, positioned almost midway between \textit{"somewhat helpful"} and \textit{"helpful"}.
This indicates that future work should look into better visualizations of the emission labels and their associated transportation modes.

\subsubsection{The popularity labels influencing their decision to avoid choosing destinations with higher popularity}

Participants were asked to rate the helpfulness of color-coded popularity labels like \textit{"rare find}, \textit{"traffic"}, \textit{"rising"}, and \textit{"hotspot"} which denote least popular, moderately popular, and most popular destinations. In~\autoref{fig: ui_labels_eval}, it is observed that 30\% of participants (approximately 60 individuals) rated these labels as \textit{"helpful"} in their decision-making, representing the largest group. Following closely, 27\% felt \textit{"very helpful"}, while 23\% found them \textit{"somewhat helpful"}. The groups identifying as \textit{"extremely helpful"} and \textit{"not helpful"} were relatively close in size, at 11\% and 9\%, respectively. 
This highlights participants' inclination to perceive popularity labels as \textit{"helpful"}, reaffirming their effectiveness in guiding users toward destinations of varying popularity and potentially steering them away from trendy locations.

\subsubsection{The seasonality labels influencing their decision to avoid choosing destinations with higher monthly seasonal demand}

This question aimed to determine if the seasonality labels ---\textit{"quiet"}, \textit{"off-peak"}, \textit{"busy"}, or \textit{"crowded"} could effectively guide users' preferences towards destinations with specific seasonality ratings, particularly influencing them to avoid crowded locations during a given month.
As evident from~\autoref{fig: ui_labels_eval}, these seasonality labels had a more significant impact on participants compared to previous questions. A majority, 33\% or roughly 66 out of 200 participants, expressed feeling \textit{"very helpful"} due to these labels. Following this, 26\% found them \textit{"helpful"}, and 20\% chose \textit{"somewhat helpful"}. A smaller fraction, 6\% or 12 participants, reported being \textit{"not helpful"} while 15\%, comprising 30 participants, found the seasonality labels as \textit{"extremely helpful"}. 
In summary, the majority found the labels \textit{"very helpful"}, with the average response closer to \textit{"helpful"}. This positive feedback highlights the value of providing seasonality labels to aid users in making informed travel decisions.

\subsubsection{The highlighting labels ("Green Recommended," "Lowest Emission," "Hidden Gem," "Least Crowded") encouraging them to choose these options over others} \label{subsection: highlighting_labels}

The purpose of this question was to assess whether these additionally highlighted labels successfully influenced users to prefer these labeled options over others.
The results indicate that the most significant proportion of respondents, 29\%, rated the highlighting labels as \textit{"somewhat helpful"} in influencing their decision-making. This group's size is nearly equivalent to those who rated it as \textit{"helpful"} or \textit{"very helpful"}, at 27\% and 28\%, respectively. Moreover, the percentage of participants who reported being \textit{"extremely helpful"} is the same as those who reported no effect, each comprising 8\% of responses. While \textit{"somewhat helpful"} was the most prevalent response, the combined responses from the \textit{"helpful"}, \textit{"very helpful"}, and \textit{"extremely helpful"} categories skew the average response towards \textit{"helpful"}.

\subsubsection{The color-coding labeling of the SF indicator influencing their decision-making process}

We color-coded the \textit{SF indicators} according to the traffic light convention, with the lowest indices in green and the highest in red. The lower values are more desirable for sustainability. Respondents were requested to rate how helpful they found this color-coded labeling system, capturing the overall sustainability of the destination and aiding their decision to select a greener alternative.
As evident from~\autoref{fig: ui_labels_eval}, a majority of respondents, 53\% or 106 out of 200, found the labels \textit{"extremely helpful"} (18\%) or \textit{"very helpful"} (35\%). Additionally, a considerable 33\% described the labels as \textit{"helpful"}. In contrast, 10\% felt the labels were only \textit{"somewhat helpful"}, and 4\% considered them \textit{"not helpful"}. In summary, \textit{"very helpful"} was the response given most frequently, with the average response falling between \textit{"helpful"} and \textit{"very helpful"}, yet leaning more towards \textit{"very helpful"}.

%% file: sections/results/3_2_results.tex
\subsection{Card View and Map View}

\begin{figure*}[htbp]
    \centering
    \begin{subfigure}{0.3\textwidth}
        \centering
        \includegraphics[width=\textwidth]{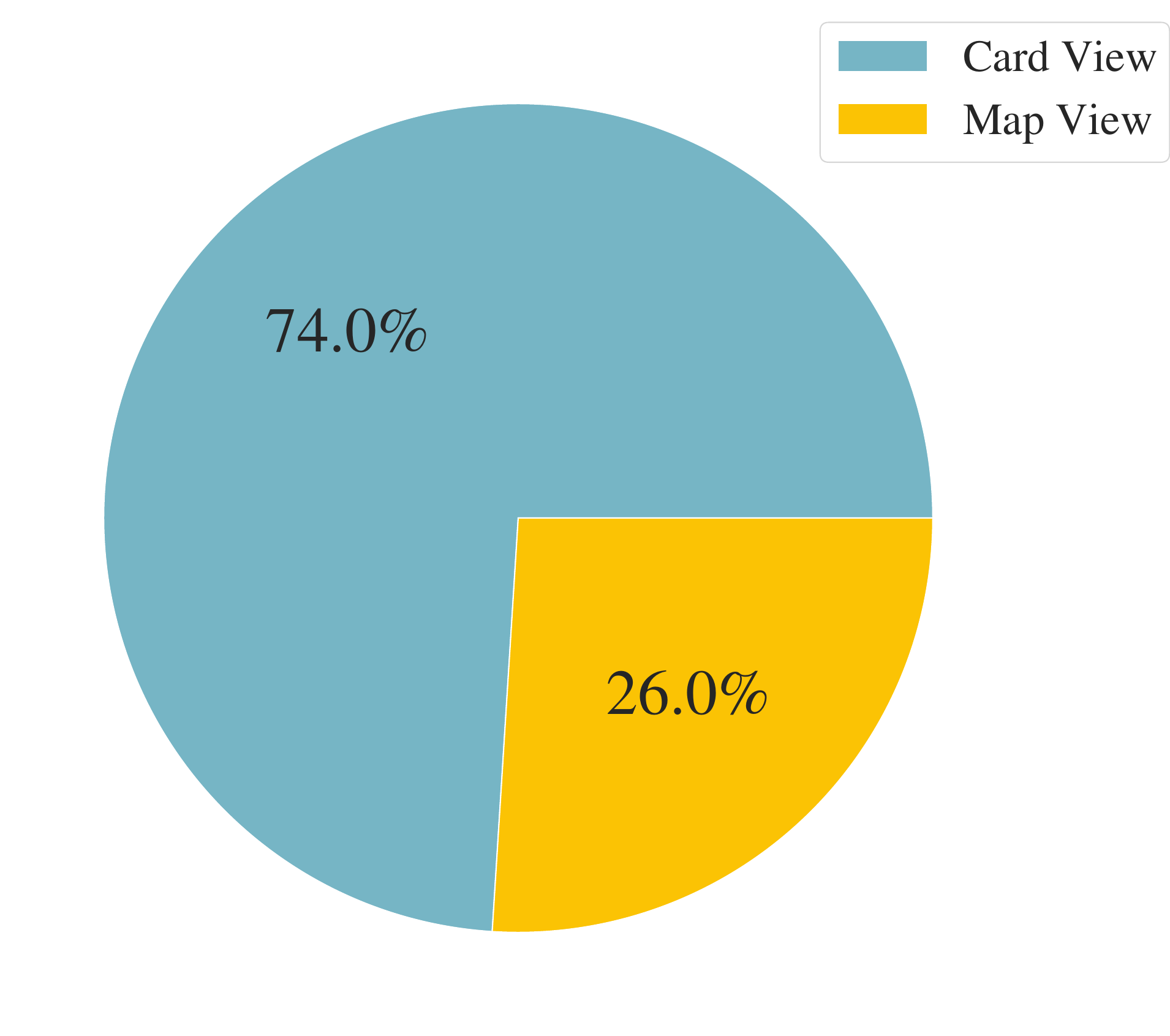}
        \caption{Map view vs Card view}
        \label{fig: map_vs_card_pie}
    \end{subfigure}
    \hfill
    \begin{subfigure}{0.65\textwidth}
        \centering
        \includegraphics[width=\textwidth]{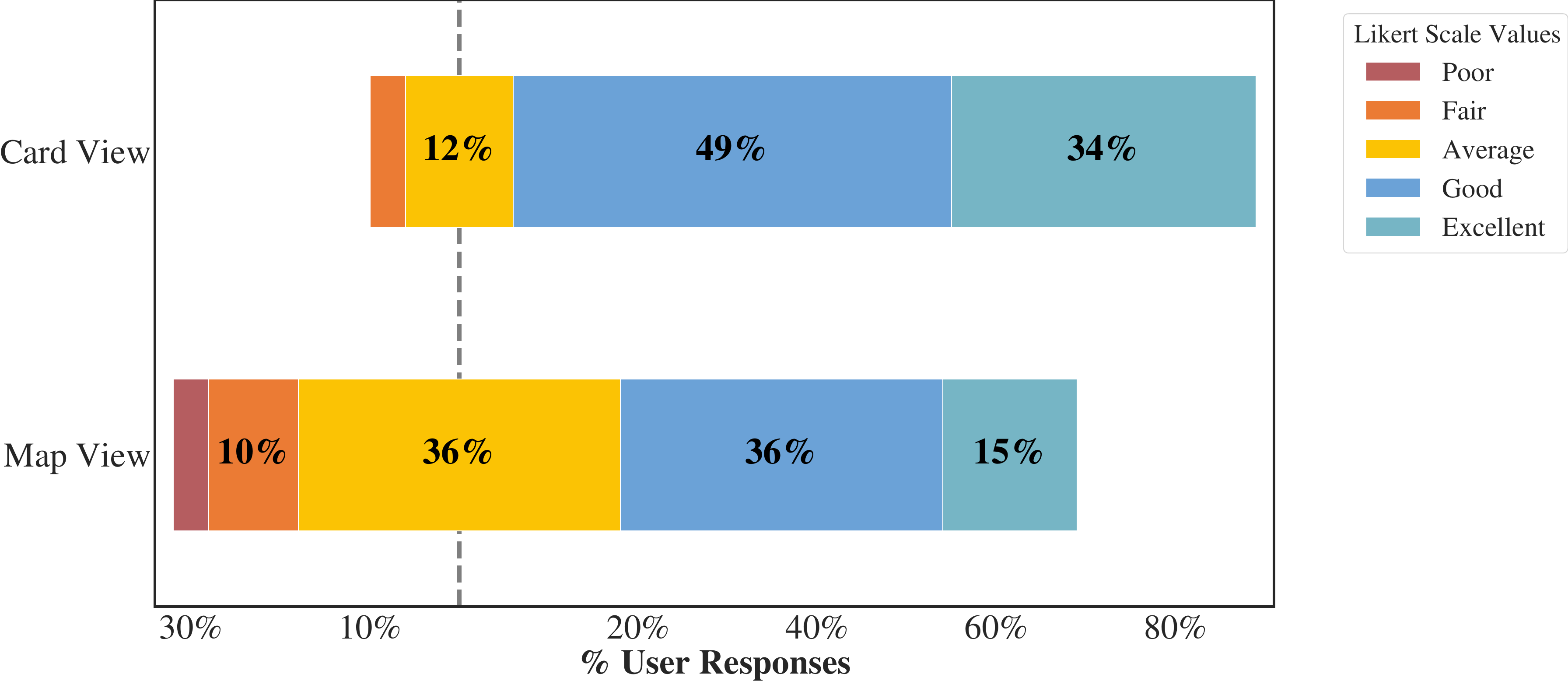}
        \caption{Map vs Card view effectiveness}
        \label{fig: map_vs_card_likert}
    \end{subfigure}
    \caption{Card View vs. Map View eﬀectiveness}
    \label{fig: card-map-effectiveness}
\end{figure*}

Users were presented with two display options: a card view and a map view, and were asked to indicate their preference. As shown in~\autoref{fig: map_vs_card_pie}, the majority (74\%) preferred the card view, with only 26\% opting for the map view. This outcome indicates a clear preference for the card view user interface, suggesting that participants found it more appealing and user-friendly compared to the map view.

~\autoref{fig: map_vs_card_likert} showcases participant evaluations of the card view and map view's effectiveness in highlighting the most sustainable travel options. 
It means that 83\% of participants (166 out of 200 users) considered the card view to be \textit{"good"} or better. A mere 12\% chose \textit{"average"} to describe its effectiveness, while a tiny fraction deemed it only \textit{"fair"} in aiding the discovery of sustainable destinations. Notably, no responses categorized the card view as \textit{"poor"}. 
On the other hand, an equal proportion of respondents, 36\% each, assessed the effectiveness of the map view as either \textit{"good"} or \textit{"average"}. Following these, 15\% rated the effectiveness as \textit{"excellent"}. A small minority, 3\%, considered the map view \textit{"poor"} in terms of effectiveness, and another 10\% described it as \textit{"fair"}. 
In conclusion, most users favored the card view over the map view, indicating a preference for the nuanced presentation of sustainability components in the user interface compared to high-level visual representations.

%% file: sections/6_conclusion.tex
\section{Conclusion} \label{section: conclusion}

This paper investigated the potential of user interface design elements to promote sustainable city trips. Using mockups and user evaluations, we explored the effectiveness of various features such as sustainability tags, emission labels, popularity indicators, seasonality information, and overall sustainability scores in influencing user behavior. 
Key findings indicate that while labels indicating popularity and seasonality successfully steered users towards less crowded and seasonally appropriate destinations, the response to emission labels associated with transportation modes was somewhat skewed, suggesting room for improvement in their visualization. The incorporation of sustainable tags alongside a color-coding scheme for the \textit{SF indicator} proved effective in conveying destination sustainability. Additionally, participants clearly preferred the card view user interface over the map view for its user-friendliness and clarity of information.

This study paves the way for future research and development in sustainable tourism applications. 
Future work aims to construct an interactive application with refined visualization methods for emissions data. 
It will incorporate real-time updates on crowd levels and traffic conditions instead of relying on mockups for evaluation. 
Furthermore, integrating user preferences and travel history could enhance personalization, resulting in more engaging recommendations.
By incorporating these findings and continuing to explore innovative design solutions, we can create user-friendly applications that empower travelers to make informed and sustainable choices, contributing to a more responsible and environmentally conscious tourism industry.